\documentclass[aps,pra,twocolumn,showpacs]{revtex4-1}

\usepackage{amsfonts,amssymb,amsmath, braket}
\usepackage{mathtools}
\usepackage[]{graphics,graphicx,epsfig}
\usepackage{amsthm,multirow}
\usepackage{color}
\usepackage{diagbox}
\begin{document}

\title{Target Detection with Asymmetric Gaussian Classical Correlation}

\author{Su-Yong Lee}
\email[]{suyong2@add.re.kr}

\author{Yonggi Jo}

\author{Jihwan Kim}

\author{Zaeill Kim}

\author{Yong Sup Ihn}

\author{Duk Y. Kim}

\affiliation{Agency for Defense Development, Daejeon 34186, Korea}

\date{\today}

\begin{abstract}
 We study the characteristics of two-mode classical correlation in Gaussian regime, which is produced by impinging a thermal state into a beam splitter. 
 It is explicitly investigated how a measurement on one mode affects the other. The result enables the distinction between classical and quantum correlations.
 Then, we show how classically correlated states are advantageous in target detection, specifically for an asymmetric classical correlation with different mean photon numbers in the signal and idler modes. Under on-off detection, the classically correlated states significantly outperform quantum correlated states as the environmental thermal mean photon number increases.
 Furthermore, we demonstrate the advantage of local Gaussian operations in the target detection.
\end{abstract}

\maketitle

\section{Introduction}
Classical correlation is a statistical behavior between two (or more) parties, where knowing one party determines the other with a probability.
It can be also represented by a self-correlation of a single-party over time.
The information shared between two (or more) parties is deterministic, which means it can be measured and duplicated without disturbing the original system. It allows for the perfect cloning of classical data bits, ensuring that classical correlation can be widely distributed and archived.
Conversely, quantum correlation which is based on the superposition principle cannot be deterministically duplicated, making the perfect cloning of unknown qunatum states physically impossible.

In Gaussian regime, quantum states are completely described with the first and second-order moments \cite{Ferraro,Wang07,Weedbrook12,Adesso} which are used to construct the Wigner distribution.
The first-order moment is represented by displacement operation, and the second-order moment is represented by covariance matrix. 
Both classical and quantum correlations are represented by the second-order moments. 
For two-mode Gaussian states, the covariance matrix is given by a four-by-four matrix, which consists of signal and idler modes.
Classical correlation is produced by impinging a thermal state into a beam splitter, where a thermal state is produced by scattering coherent states with a rotating ground glass disk or by blocking one mode of a two-mode squeezed vacuum (TMSV) state. 
On the other hand, quantum correlations including entanglement and quantum discord, are produced by nonlinear process, such as parametric down conversion and Kerr effect. TMSV state is a representative of Gaussian quantum correlation.

A target detection is on the purpose of discriminating two different states in a noisy environment, such as the presence and absence of a target \cite{Lloyd}. 
The presence of the target is characterized by a state that interacted weakly with the target, whereas its absence is characterized by a state completely unaffected by the target.
For Gaussian states, TMSV state presents the best performance in the target detection protocol \cite{Tan}.
A classically correlated state cannot beat the TMSV state but can approach the performance of a coherent state \cite{Lee} which is the standard quantum state. There are several different measurement setups \cite{Lee,Guha,Zhuang,Jo,Yang,Lee23,Reichert,Shi,Jeon} proposed for target detection with Gaussian states.
The study of target detection was extended to include the effects of both Gaussian \cite{Jo21,Kim23} and non-Gaussian \cite{Zubairy,Zhong,Palma} operations on Gaussian states.

Here we study the characteristics of Gaussian classical correlation, compared to Gaussian quantum correlation.
Then, we investigate the asymmetric Gaussian classical correlation in target detection with on-off detection, resulting in beating the performance of quantum correlation. 
Note that it is not possible to study asymmetric Gaussian quantum correlation with TMSV state. But there is a study of asymmetric non-Gaussian quantum correlation for target sensing, using entangled coherent states \cite{Lee24} which are flexible to control the ratio of the mean photon numbers between the signal and idler modes.
Moreover, we consider local Gaussian operations on the signal mode, such as displacement operation and single-mode squeezing operation, and demonstrate which one is more advantageous in a target detection with classical correlation. 


\section{Characteristics of Classical correlation in Gaussian regime}

\begin{figure}
\centering
\includegraphics[width=0.45\textwidth]{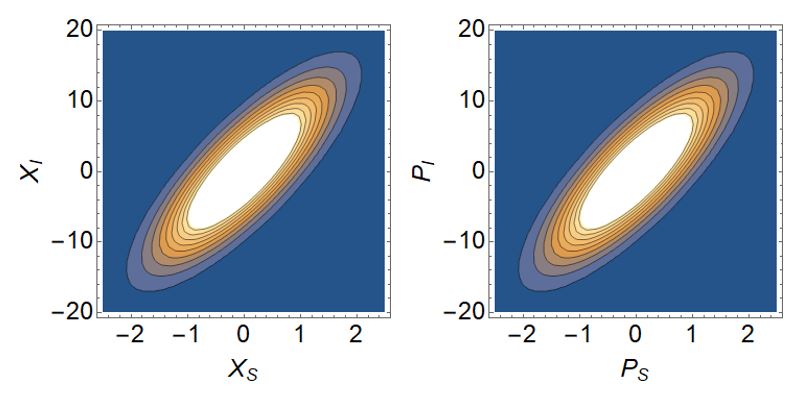}
\caption{Gaussian classical correlation in phase spaces of $(X_S,X_I)$ and $(P_S,P_I)$, at $N_S=1,~N_I=100$.
It always presents the same orientation in the correlation phase spaces. }
\label{classical}
\end{figure}

A thermal state is injected into one of the input ports of a beam splitter, where the transmittance(reflectance) determines the signal(idler) mean photon number.
The output two-mode classically correlated state is described with a four-by-four covariance matrix 
 $V_{\text{SI}}=\langle [\hat{a}_S~ \hat{a}_I~ \hat{a}_S^{\dag}~ \hat{a}_I^{\dag}]^T[ \hat{a}_S^{\dag}~ \hat{a}_I^{\dag}~ \hat{a}_S~ \hat{a}_I]\rangle$, where the subscript $S$($I$) represents the signal(idler) mode and $\hat{a}_{k}$ is the annihilation operator in mode $k(=S,I)$ that is also described with position and momentum operators 
 $\hat{X}_k$ and $\hat{P}_k$. Its explicit formula is given by
\begin{align}
	 V_{\text{SI}}=\begin{pmatrix}
 N_S+1 & \sqrt{N_S N_I} & 0 & 0 \\
\sqrt{N_S N_I} & N_I+1 & 0 & 0 \\
0 & 0 & N_S & \sqrt{N_S N_I} \\
0 & 0 &\sqrt{N_S N_I} & N_I
	\end{pmatrix},
\end{align}
where $N_{S(I)}$ is the mean photon number of the signal (idler) mode, and 
$\langle \hat{a}_S^{\dag}\hat{a}_I \rangle=\langle \hat{a}_I^{\dag}\hat{a}_S \rangle=\sqrt{N_S N_I}$ is the classical correlation term.
Fig.~\ref{classical} represents the Gaussian classical correlation which exhibits the same orientation in both the $(X_S,X_I)$ and $(P_S,P_I)$ spaces.
From now on, it is compared with a TMSV state that exhibits orthogonal orientations between $(X_S,X_I)$ and $(P_S,P_I)$ spaces.
Tracing out the idler mode of the classically correlated state, the signal mode is changed into a thermal state with $N_S$, 
which is identical to tracing out the idler mode of the TMSV state. 
It is impossible to discriminate between the two states after tracing out one mode.

We investigate how a measurement on one mode influences the other mode, where classically correlated state of Eq. (1) presents quite different results from the TMSV state.
After performing a measurement on the idler mode, the signal mode is changed into a state, resulting in Gaussian or non-Gaussian, and classical or non-classical states. We consider an on-off detection, a photon number resolving (PNR) detection, and a homodyne detection. The on-off (or PNR) detection changes a Gaussian state into a non-Gaussian state, but the homodyne detection does not.

First, after performing on-off detection on the idler mode and post-selecting only the click events, the classically correlated state is transformed into 
\begin{eqnarray}
\rho_{S}^{(on)}=\frac{1+N_I}{N_I}\sum^{\infty}_{n=0}\bigg[\frac{N^n_S}{(1+N_S)^{n+1}}-\frac{N^n_S}{(1+N_T)^{n+1}}\bigg]|n\rangle_S\langle n|,\nonumber\\
\end{eqnarray}
where $N_T=N_S+N_I$. It is a classical state that exhibits $\frac{3}{2}<g^{(2)}(0)<2$, which is confirmed by the degree of second-order coherence for zero-time delay $1\leq g^{(2)}(0)$ \cite{Loudon}. Moreover, its Glauber-Sudarshan P-function \cite{Zubairy} presents non-negative values.
Correspondingly, TMSV state is transformed into a vacuum-truncated thermal state, $\frac{1+N_S}{N_S}\sum^{\infty}_{n=1}\frac{N^n_S}{(1+N_S)^{n+1}}|n\rangle_S\langle n|$ which is a non-classical state exhibiting $0<g^{(2)}(0)<2$ and negative values of the P-function.

Second, after performing PNR detection on the idler mode, the classically correlated state is transformed into 
\begin{eqnarray}
\rho_S^{(k)}=\frac{(\frac{1+N_I}{1+N_T})^{1+k}}{k!}\sum^{\infty}_{n=0}\frac{(n+k)!}{n!}(\frac{N_S}{1+N_T})^n |n\rangle_S\langle n|, 
\end{eqnarray}
where $k$ is the number of photons detected in the idler mode.
It is a classical state that exhibits $1<g^{(2)}(0)=\frac{2+k}{1+k}<2$ and non-negative values of the P function.
TMSV state is transformed into a Fock state $|k\rangle$ which is the most non-classical state exhibiting $g^{(2)}(0)=1-\frac{1}{k}$ and negative values of the P-function.

Third, after performing homodyne detection $|X\rangle_I\langle X|$ on the idler mode, the classically correlated state is transformed into a displaced thermal state. 
Here we describe it with a Wigner function,
\begin{eqnarray}
W(X,P)=\frac{1}{2\pi\sqrt{\sigma^2_X\sigma^2_P}}\exp\bigg[-\frac{(X-\langle \hat{X}\rangle)^2}{2\sigma^2_X}-\frac{P^2}{2\sigma^2_P}\bigg],\nonumber\\
\end{eqnarray}
where $\sigma^2_X=\frac{1}{2}(1+\frac{2N_S}{1+2N_I})$, $\sigma^2_P=\frac{1}{2}(1+2N_I)$, and $\langle \hat{X}\rangle=\frac{\sqrt{N_S N_I}}{N_I+1/2}$.
The uncertainty of position and momentum is given by $\sigma^2_X \sigma^2_P=\frac{1}{4}(1+2N_S+2N_I)$ which is greater than the minimum uncertainty of $\frac{1}{4}$.
The displaced thermal is a classical state with $g^{(2)}(0)\geq 1$ and non-negative values of the P-function.
TMSV state is transformed into a displaced squeezed state which has 
$\sigma^2_X =\frac{1}{2\cosh(2r)},~\sigma^2_P=\frac{\cosh(2r)}{2},~\langle \hat{X}\rangle=\tanh(2r)$. 
It has a minimum uncertainty of $\sigma^2_X \sigma^2_P=\frac{1}{4}$. It is a non-classical state that can exhibit $g^{(2)}(0)<1$ and negative values of the P-function.

Moreover, we may consider double homodyne detection which is also represented by a projection of a coherent state basis of 
$|\beta\rangle=|\frac{X+iP}{\sqrt{2}}\rangle$. After performing double homodyne detection $|\beta\rangle_I \langle \beta|$ on the idler mode,
the classically correlated state is transformed into a displaced thermal state which is given with a Wigner distribution,
\begin{eqnarray}
W(X,P)=\frac{1}{2\pi \sigma^2}\exp\bigg[-\frac{(X-\langle \hat{X}\rangle)^2+(P-\langle \hat{P}\rangle)^2}{2\sigma^2}\bigg],\nonumber\\
\end{eqnarray}
where $\sigma^2=\frac{1}{2}(1+\frac{2N_S}{1+N_I}),~\langle \hat{X}\rangle=\langle \hat{P}\rangle=\frac{\sqrt{N_S N_I}}{N_I+1}$. The uncertainty of position and momentum is greater than $\frac{1}{4}$.
Correspondingly, TMSV state is transformed into a coherent state, such as $|\sqrt{\frac{N_S}{1+N_S}}\beta\rangle$ having a minimum uncertainty of $\sigma^2=\frac{1}{2}$.
It is the standard quantum state with $g^{(2)}(0)=1$.

The above measurements on one mode can produce non-classical properties on the other mode, not for classically correlated state but for TMSV state.
Thus, one can discriminate between classical and quantum correlations by performing a measurement on one mode and observing the conditional state of the other mode.

\section{Target detection}
The performance of the target detection is limited by quantum Chernoff bound (QCB) \cite{QCB}. 
For Gaussian states, the TMSV state outperforms the classically correlated state by a factor of four in the error exponent \cite{Tan,Lee}.
Under current technology, the TMSV state cannot achieve the optimal bound since it requires collective measurements with a quantum memory \cite{Tan,Shi}.
However, the classically correlated state can achieve its optimal bound with a photon-number-difference measurement after mixing the reflected signal and the remaining idler modes at a 50:50 beam splitter \cite{Lee}.
We consider simple measurement setups to compare the performance of the classically correlated state and the TMSV state.

After the classically correlated state reflects off a low-reflectance target, the output state is described with the covariance matrix  
$V_{RI}^{(jk)}=\langle \hat{R}_j\hat{R}_k+\hat{R}_k\hat{R}_j\rangle-2\langle \hat{R}_j\rangle\langle \hat{R}_k\rangle$,
where $\hat{R}_k=\hat{X}_k~(\hat{P}_k)$ and $\hat{a}_k=\frac{1}{\sqrt{2}}(\hat{X}_k+i\hat{P}_k)$.
Its explicit formula is given by
\begin{align}
	 V_{\text{RI}}=\begin{pmatrix}
 2\langle \hat{X}^2_R\rangle & 0 & 2\langle \hat{X}_R\hat{X}_I\rangle & 0 \\
0 & 2\langle \hat{P}^2_R\rangle & 0 & 2\langle \hat{P}_R\hat{P}_I\rangle \\
2\langle \hat{X}_R\hat{X}_I\rangle & 0 & 1+2N_I & 0 \\
0 & 2\langle \hat{P}_R\hat{P}_I\rangle & 0 & 1+2N_I
	\end{pmatrix},
\end{align}
where $\langle \hat{X}^2_R\rangle=\langle \hat{P}^2_R\rangle=\frac{1}{2}(1+2N_B+2\kappa N_S)$ and $\langle \hat{X}_R\hat{X}_I\rangle=\langle \hat{P}_R\hat{P}_I\rangle=\sqrt{\kappa N_S N_I}$. The subscript $R$ represents the reflected mode, and $\kappa$ is the target reflectance.

\subsection{On-off detection}

Given a measurement setup, we can evaluate the performance with the signal-to-noise ratio (SNR) that is the exponent of the target detection error rate 
$P_{\text{err}}=\frac{1}{2}\exp[-M\times \text{SNR}]$.
The SNR is defined as $(\langle \hat{O}\rangle_{\kappa}-\langle \hat{O}\rangle_{\kappa=0})^2/2(\sqrt{\Delta^2 O_{\kappa}}+\sqrt{\Delta^2 O_{\kappa=0}})^2$,
where $\hat{O}$ is an observable.
Under a homodyne detection, the classically correlated state can approach the performance of the TMSV state \cite{Lee}.
Under a PNR detection, the classically correlated state cannot outperform the TMSV state \cite{Lee23}.

\begin{figure}
\centering
\includegraphics[width=0.45\textwidth]{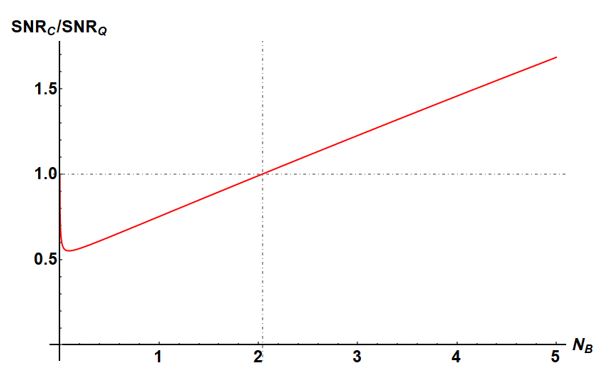}
\caption{SNR ratio as a function of $N_B$ at $N_S=1,~N_I=100$.
$\text{SNR}_{\text{C}}$ is for the classically correlated state and $\text{SNR}_{\text{Q}}$ is for the TMSV state in the target detection with on-off detection.
The crossing point of  $\text{SNR}_\text{C}/\text{SNR}_\text{Q}=1$ is at $N_B\approx2.04$.
As $N_B \rightarrow 0$, the SNR ratio tends to unity.
}
\label{SNR}
\end{figure}

We consider the most feasible measurement setup, such as on-off detection which extracts less information from classical and quantum correlation, compared to PNR detection and homodyne detection.
The observable for coincidence `ON' detection is given by $\hat{O}_{\text{on}}=(\hat{I}_R-|0\rangle_R\langle 0|)\otimes (\hat{I}_I-|0\rangle_I\langle 0|)$,
where $\hat{I}$ is the identity operator and $|0\rangle\langle 0|$ is the vacuum operator.
It has been partially investigated in \cite{Lee23}.
Given a fixed $N_S$ and $N_S<N_I$, the classically correlated state can outperform the TMSV state under on-off detection, 
with an increase in the mean thermal photon number $N_B$. At $\kappa N_S\ll1$, each SNR is given by
\begin{eqnarray}
\text{SNR}_{\text{C}}&\approx&\frac{\kappa^2 N_S^2 N_I (2+N_I)^2}{8N_B(1+N_I+N_B)(1+N_I)^2(1+N_B)^2},\nonumber\\
\text{SNR}_{\text{Q}}&\approx&\frac{\kappa^2 N_S (1+N_S)^2}{8N_B(1+N_S+N_B)(1+N_B)^2}.
\end{eqnarray}
At $\kappa N_S\ll 1 <N_S N_I$, the threshold of the mean thermal photon number for the classically correlated state to outperform the TMSV state is given by
\begin{eqnarray}
N_B^{\text{th}} \equiv \frac{(N_I-N_S)(1+N_S)(1+N_I)}{(1+N_I)^2 N_S-(1+N_S)^2 N_I}.
\end{eqnarray}
When $\kappa=0.01$, $N_S=1$, and $N_I=100$, the  threshold of the mean thermal photon number is approximately $2.04$.
The SNR ratio is given as a function of the mean thermal photon number $N_B$ in Fig.~\ref{SNR}.  
As $N_B\rightarrow \infty$, the SNR of the classically correlated state surpasses that of the TMSV state by a factor of up to $25.5$.
When $\kappa=0.01$, $N_S=0.001$, and $N_I=10^5$, the  threshold of the mean thermal photon number is approximately $10^3$.
As $N_S$ decreases, the threshold of the mean thermal photon number increases. 

\subsection{Local Gaussian operation}

The optimal performance of the target detection error rate is given as $P_{err}\sim \exp(-M\gamma)$, where $\gamma$ represents the QCB.
We assume that the target reflectance is much smaller than unity, such that the simplified formula \cite{Kim23} can be applied to the following scenarios:
Displacement operation and single-mode squeezing operation on the signal mode of the classically correlated state before interacting with a target.

After applying the displacement operation $\hat{D}(\alpha)=\exp(\alpha\hat{a}^{\dag}-\alpha^{\ast}\hat{a})$ to the signal mode, its signal mean photon number is given by $\langle \hat{N}_S\rangle=N_S+|\alpha|^2$.
Then, after interaction with the target, the output covariance matrix of Eq. (6) remains unchanged but the first-order moment is added to the covariance matrix, which results in displacement $\bold{d}^T=\sqrt{2\kappa}(|\alpha|\cos\phi, |\alpha|\sin\phi)$, where $\alpha=|\alpha|e^{i\phi}$. 
The corresponding optimal performance $\gamma_{\text{dis}}$ is given by
\begin{eqnarray}
\gamma_{\text{dis}}=\kappa && \bigg[ \frac{ N_S N_I}{(\sqrt{N_B(1+N_I)}+\sqrt{(1+N_B)N_I})^2}\nonumber\\
&&+\frac{ |\alpha|^2}{(\sqrt{N_B}+\sqrt{1+N_B})^2} \bigg],
\end{eqnarray}
where the second term represents the effect of the displacement operation, which is independent of the classical correlation.
The performance of classical correlation is enhanced by the separate term of displacement operation.

\begin{figure}
\centering
\includegraphics[width=0.45\textwidth]{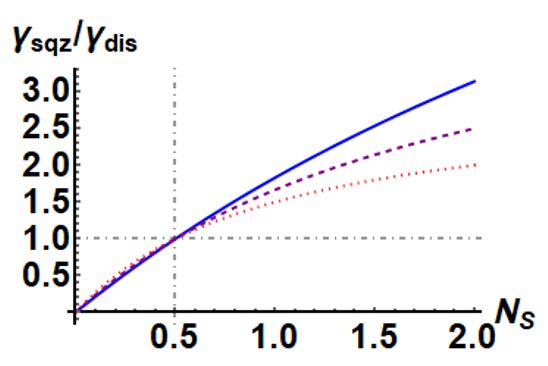}
\caption{Optimal performance ratio ($\gamma_{\text{sqz}}/\gamma_{\text{dis}}$) as a function of $N_S$ at $N_B=600,~N_I=100,~\kappa=0.01$.
$|\alpha|^2=\sinh^2{r}=$ $1$(red dotted), $2$(purple dashed), $5$(blue solid).
}
\label{sqz_dis}
\end{figure}

\begin{figure}
\centering
\includegraphics[width=0.45\textwidth]{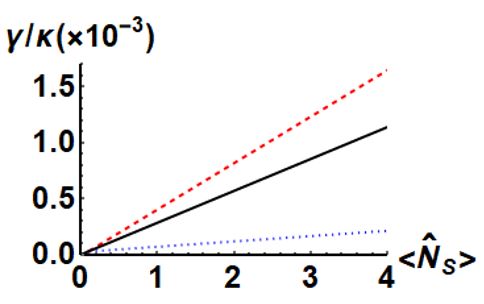}
\caption{Comparison among $\gamma,~\gamma_{\text{sqz}}, ~\gamma_{\text{dis}}$ as a function of $\langle \hat{N}_S\rangle $ at $N_B=600,~N_I=1,~N_S=0.1$.
Displaced operation (red dashed), single-mode squeezing operation (blue dotted), and no operation (black solid) on the classically correlated state.
}
\label{comparison}
\end{figure}

After applying the single-mode squeezing operation $\hat{S}(r)=\exp[\frac{r}{2}(\hat{a}^2-\hat{a}^{\dag 2})]$ to the signal mode, 
its signal mean photon number is given by $\langle \hat{N}_S\rangle=N_S+\sinh^2{r}+2N_S\sinh^2{r}$.
Then, after interaction with the target, the covariance matrix elements of Eq. (6) are changed as follows: 
$\langle \hat{X}^2_R\rangle=\frac{1}{2}[1-\kappa+2N_B+\kappa(1+2N_S)e^{-2r}]$, 
$\langle \hat{P}^2_R\rangle=\frac{1}{2}[1-\kappa+2N_B+\kappa(1+2N_S)e^{2r}]$,
$\langle \hat{X}_R\hat{X}_I\rangle=e^{-r}\sqrt{\kappa N_S N_I}$, and  
$\langle \hat{P}_R\hat{P}_I\rangle=e^{r}\sqrt{\kappa N_S N_I}$.
The corresponding optimal performance $\gamma_{\text{sqz}}$ is given by
\begin{eqnarray}
\gamma_{\text{sqz}}=\kappa N_S N_I && \bigg[ \frac{ \sinh^2{r}}{(\sqrt{N_B N_I}+\sqrt{(1+N_B)(1+N_I)})^2}\\
&&+\frac{ 1+\sinh^2{r}}{(\sqrt{N_B(1+N_I)}+\sqrt{(1+N_B)N_I})^2} \bigg],\nonumber
\end{eqnarray}
where the effect of the single-mode squeezing operation amplifies the classical correlation.
The performance of the classical correlation is scaled by a factor derived from the single-mode squeezing operation.

As $\kappa\ll 1 \ll N_I, N_B$ and $|\alpha|^2=\sinh^2{r}$, the $\gamma_{\text{sqz}}$ outperforms the $\gamma_{\text{dis}}$ at $N_S >1/2$, 
as shown in Fig.~\ref{sqz_dis}. The optimal performance ratio ($\gamma_{\text{sqz}}/\gamma_{\text{dis}}$) increases as $|\alpha|^2=\sinh^2{r}$ increases.
For a fixed $\langle \hat{N}_S\rangle $,
the displaced classically correlated state can outperform the original classically correlated state whereas the performance of the squeezed classically correlated state is inferior to that of others, as shown in Fig.~\ref{comparison}.
As $N_I$ increases, the displaced classically correlated state converges to the performance of the original classically correlated state.

\section{Summary and Discussion}
We investigated the characteristics of the asymmetric Gaussian classical correlation, where measuring one mode and observing the other allows us to discriminate classical correlation from quantum correlation. Using classically correlated states in target detection, we showed that classically correlated states can significantly outperform quantum correlated states under on-off detection. Applying local Gaussian operation on a signal mode of the classically correlated state, such as displacement operation and single-mode squeezing operation, we demonstrated that the effect of the squeezing operation is more advantageous than the displacement operation as the initial signal mean photon number ($N_S$) increases. For a fixed $\langle \hat{N}_S\rangle$, the signal mean photon number after the local Gaussian operation on the signal mode, we obtained that displaced classically correlated state can beat squeezed classically correlated state as well as the original classically correlated state.

While fundamentally classical correlation cannot replace quantum correlation, the former can outperform the latter in the target detection with on-off detection.
Depending on the measurement setup, classical correlation has potential to outperform quantum correlation in a specific protocol.
We also emphasize that Gaussian classical correlations demonstrate quite different results from Gaussian quantum correlations in local Gaussian operation on the signal mode. 
For quantum correlation represented by the TMSV state, displacement operation provides better performance than the single-mode squeezing operation \cite{Kim23}.
For a fixed $\langle \hat{N}_S\rangle $, the displaced or squeezed TMSV state cannot outperform the TMSV state \cite{Kim23}.
As a further work, we may consider non-Gaussian operations on the signal mode of the classically correlated state in order to figure out how it is useful in target detection.

\begin{acknowledgments}
This work was supported by the Institute of Information \& Communications Technology Planning \& Evaluation(IITP) and Defense Acquisition Program Administration(DAPA) grant funded by the Korea government(MSIT)  (No.2026-25607569, Integrated Fiber-FSO Quantum-Secured Communication Technology for MUM-T (FSO)).

\end{acknowledgments}

\end{document}